# MORPES: A MODEL FOR PERSONALIZED RENDERING OF WEB CONTENT ON MOBILE DEVICES

K.S.Kuppusamy[1] and G.Aghila[2]

[1]Department of Computer Science, School of Engineering and Technology, Pondicherry University, Pondicherry, India
kskuppu@gmail.com

[2]Department of Computer Science, School of Engineering and Technology, Pondicherry University, Pondicherry, India
aghilaa@yahoo.com

## ABSTRACT

*With the tremendous growth in the information communication sector, the mobile phones have become the prime information communication devices. The convergence of traditional telephony with the modern web enabled communication in the mobile devices has made the communication much effective and simpler. As mobile phones are becoming the crucial source of accessing the contents of the World Wide Web which was originally designed for personal computers, has opened up a new challenge of accommodating the web contents in to the smaller mobile devices. This paper proposes an approach towards building a model for rendering the web pages in mobile devices. The proposed model is based on a multi-dimensional web page segment evaluation model. The incorporation of personalization in the proposed model makes the rendering user-centric. The proposed model is validated with a prototype implementation.*

## 1. INTRODUCTION

The usage of mobile devices is on a steep raise all across the globe. A recent study declares that there exist around 5.9 billion mobile subscribers which accounts for 87% of the total world's population [1]. In parallel with the mobile devices usage, web access on mobile phones is also on a massive increment. The study [1] figures out that there exist around 1.2 billion mobile web users. More than 80% of existing mobile devices are capable of accessing the web content.

The World Wide Web was originally designed for personal computers which obviously have larger screens than the mobile devices. Hence, displaying the contents of web pages on mobile devices which has smaller screens has emerged as an important research problem to solve. With mobile devices taking the centre-stage of web access this problem becomes more important to solve [2]. Accessing the contents of World Wide Web on mobiles has its own unique requirements [3]. A detailed survey on how the web is accessed in mobile devices is discussed in [4]. The study in [4] explores various categories of activities carried out on mobile web access.

This paper proposes a model titled "MORPES" (MObile Rendering of Pages using Segment Evaluation). The objectives of the proposed MORPES model are as listed below:

International Journal in Foundations of Computer Science & Technology (IJFCST),Vol. 2, No.2, March 2012

- Proposing a model for rendering the web pages using a multi-dimensional segment evaluation model.

- Incorporating personalization in to the rendering process to provide a user-centric mobile web experience.

The remainder of this paper is organized as follows: In Section 2, some of the related works carried out in this domain are explored. Section 3 deals with the proposed model's mathematical representation and algorithms. Section 4 is about prototype implementation and experiments. Section 5 focuses on the conclusions and future directions for this research work.

## 2. RELATED WORKS

This section walks though some of the related works which have been carried out in this domain. The proposed MORPES model incorporates the following two major fields of study:

- Mobile Web Access

- Web Page Segmentation

### 2.1 Mobile Web Access

Accessing the contents of World Wide Web through mobile devices is inherently challenging due to the smaller sized screens used in the mobile devices. Rendering of web pages in mobile devices is carried out using various approaches.

An approach towards displaying the web page on the mobile devices is by providing an image thumbnail of page and based on the user interaction zooming the page in to relevant areas [5][6][7]. This approach requires the user to select the particular area and zoom-in to view the contents.

Another approach to render the web pages on mobile devices is to change the layout structure of the page to fit in the mobile display. A similar approach is followed in many popular mobile based browsers like Opera [8]. The problem with this approach is that for many web pages it adds a longer vertical scrollbar which makes it hard for the user to locate the information which is at the end of the page.

Modifying the structure of the web page is another approach to render the contents in a mobile device. In this approach portions of the web page are rendered at a time [9] [10][11]. An approach to filter out user specific portions of web page to display them in a mobile display in explored in [12]. Snap2Read explores an approach to convert the snaps of the magazine in to a readable form in mobile devices [13]. A web page segmentation based approach to display the contents on a mobile device is explored in [14].

The approach followed in the proposed MORPES model utilizes the segmentation of web pages. The proposed model incorporates the segment evaluation process in displaying the segments.

### 2.2 Web Page Segmentation

Web page segmentation is an active research topic in the information retrieval domain in which a wide range of experiments are conducted. Web page segmentation is the process of dividing a web page into smaller units based on various criteria. The following are four basic types of web page segmentation methods. They are Fixed length page segmentation, DOM based page segmentation, Vision based page segmentation and Combined / Hybrid method

A comparative study among all these four types of segmentation is illustrated in [15]. Each of above mentioned segmentation methods have been studied in detail in the literature. Fixed length page segmentation is simple and less complex in terms of implementation but the major problem with this approach is that it doesn't consider any semantics of the page while segmenting. In DOM base page segmentation, the HTML tag tree's Document Object Model would be used while segmenting. An arbitrary passages based approach is given in [16]. Vision based page segmentation (VIPS) is in parallel lines with the way, humans views a page. VIPS [17] is a popular segmentation algorithm which segments a page based on various visual features.

Apart from the above mentioned segmentation methods a few novel approaches have been evolved during the last few years. An image processing based segmentation approach is illustrated in [18]. The segmentation process based text density of the contents is explained in [19]. The graph theory based approach to segmentation is presented in [20].

## 3. THE MODEL

This section explores about the building blocks of propose MORPES model for rendering of web page contents in mobile displays in a personalized manner. The block diagram of MORPES model is as shown in Fig. 1.

The proposed MORPES model follows a proxy server based approach. All the page request from the mobile device would be routed to a proxy server. The proxy server fetches the pages and modifies it according to the needs of the mobile display. The proposed approach has the advantage of saving the battery power at the mobile unit. If all the restructuring happens at the mobile device then it would consume a considerable amount of battery. In order to avoid this, the proposed approach follows the proxy server based approach.

The components of the proposed model are as explained below:

- Page Fetcher: This component's role is to fetch the contents of the source page for further processing.

- Segmentor: The Segmentor receives the source page as input and segments it in to a collection. The proposed model employs Document Object Model based Segmentation.

- Segment Evaluator: The segment evaluator's role is to evaluate the segment weights. The proposed model employs a variation of the MUSEUM (Multi dimensional Segment Evaluation Model) [21]. This model evaluates the segments by considering various structural semantics.

- Profile Builder: The profile builder component's role is to build the profile of the user in the form of keyword vectors.

- Profile Store: The profile store is a repository of all the profile terms gathered by the profile builder.

- Profile Updator: The profile updater would incrementally update the profile created by the user. The profile updater monitors the activities of the users and updates the profile accordingly.

- Template Manager: The template manager holds various customized templates used to render the selected segments.

International Journal in Foundations of Computer Science & Technology (IJFCST),Vol. 2, No.2, March 2012

- Score Sorter: The score sorter would sort the segments based on the scores calculated by the evaluator.
- Renderer: The renderer component renders the segments using the template supplied by the template manager.
- Segment Buffer: The segment buffer would hold the segments with the top score's to be displayed next.

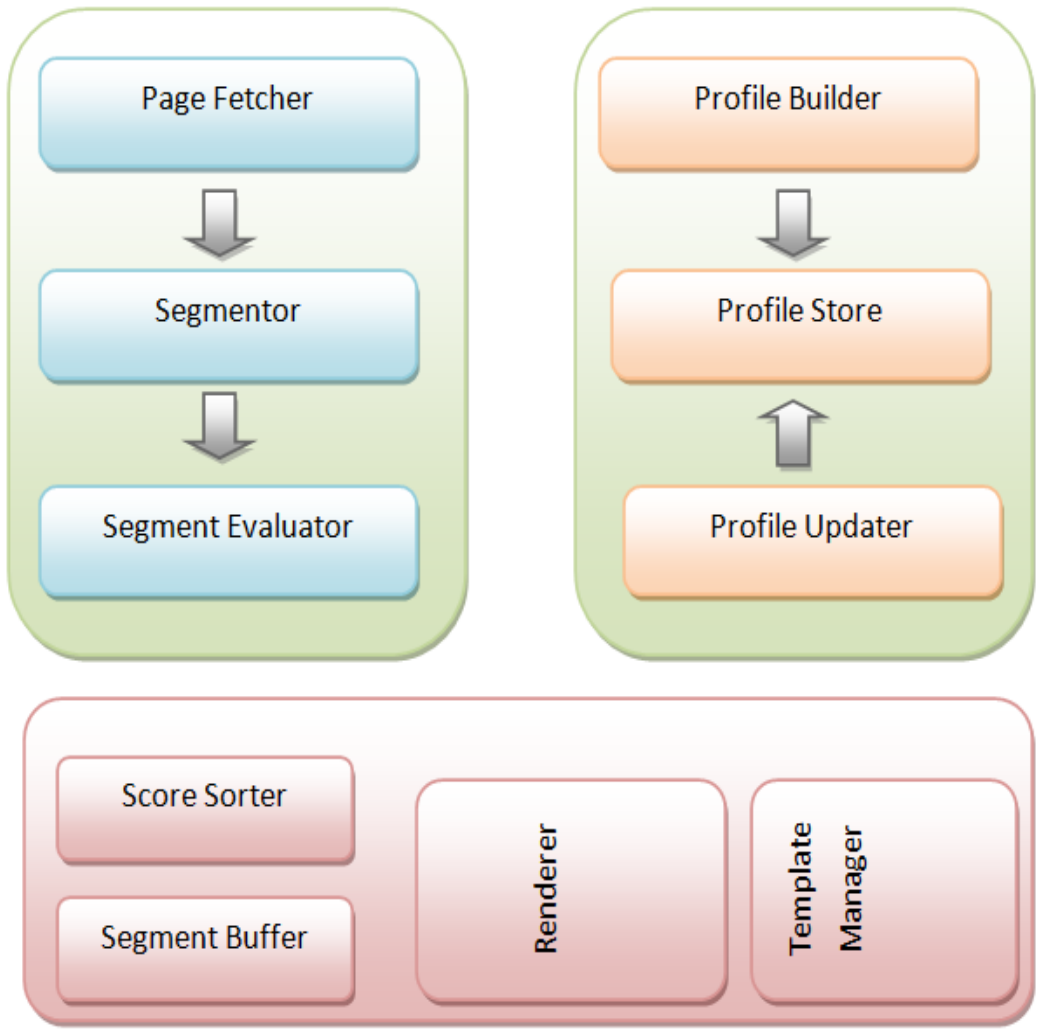

Figure 1.  Block Diagram of MORPES Model

### 3.1 Mathematical Model

The proposed MORPES model splits the source page into various segments.

$$\Omega = \{\omega_1, \omega_2, \omega_3 ... \omega_n\} \tag{1}$$

The user profile terms are stored in the profile store by the profile builder.

International Journal in Foundations of Computer Science & Technology (IJFCST),Vol. 2, No.2, March 2012

$$\Gamma = \{\alpha_1, \alpha_2, \alpha_3 ... \alpha_n\} \quad (2)$$

In (2), each $\alpha_i$ represents a profile term corresponding to the user. Each of the segments identified in (1), would be evaluated to yield a score. The evaluation is done using a variation of the MUSEUM model. Each segment is evaluated using various structural semantics of the segment.

$$\Phi = \{\forall_{i=1..n} \delta(\omega_i, \Gamma)\} \quad (3)$$

In (3), the function $\delta(\omega_i, \Gamma)$ indicates the calculation of segment score. The segment score is calculated using five different dimensions as shown in (4).

$$\delta(\omega_i, \Gamma) = \begin{cases} \forall L : \mu(L_i, \Gamma) & \oplus \\ \forall I : \mu(I_i, \Gamma) & \oplus \\ \forall T : \mu(T_i, \Gamma) & \oplus \\ \forall V : \mu(V_i, \Gamma) & \oplus \\ \forall F : \mu(F_i, \Gamma) & \end{cases} \quad (4)$$

In (4), $L$ indicates the links, $I$ indicates the Images, $T$ indicates theme of the page, $V$ indicates the visual features and $F$ indicates the freshness weight. The function $\mu$ indicates the score calculation with respect to the corresponding dimension. The symbol $\oplus$ in each row in (4) indicates the summation of scores.

The segment sorter would sort the segments based on the weights calculated in (4) as shown in (5).

$$\Psi = \{\forall_{i=1..n} \langle \omega_i, \delta_i \rangle : \delta_{i-1} > \delta_i\} \quad (5)$$

The template manager would hold various templates and a particular template identified by the user or by the system would be used to render the initial segments.

$$\Pi = \{\lambda_1, \lambda_2, \lambda_3 ... \lambda_n\} \quad (6)$$

The segments in $\Psi$ are selected based on their score. The renderer module displays the segments based on the template selected by the template manager and top scoring segments in $\Psi$, in the first shot.

$$\Upsilon_i = \Psi(\omega_i) / \vec{\lambda} \quad (7)$$

In (7), $\Upsilon_i$ indicates the group of segments displayed in the shot "i". The $\vec{\lambda}$ indicates the selected template. The low scoring segment would be rendered in the successive shots.

The proposed model renders the web pages in to small screen mobile devices by breaking the page into various segments. The speciality of the model is the multidimensional score calculation using structural semantics of the page which incorporates the user specific requirements. This makes the rendering specific to the requirements of the user. So two users, visiting the same web page may get different set of segments at the initial shot, depending on their profile terms.

International Journal in Foundations of Computer Science & Technology (IJFCST),Vol. 2, No.2, March 2012

## 3.2 The Algorithm

The algorithmic representation of the proposed MORPES model is depicted in this section.

Algorithm MORPES

Input: Source Web Page $\Omega$ , profile $\Gamma$

Output : segments for mobile to render

Begin

1. Split the source page into various segment

   $\Omega = \{\omega_1, \omega_2, \omega_3 \ldots \omega_n\}$

2. Fetch the user profile terms

   $\Gamma = \{\alpha_1, \alpha_2, \alpha_3 \ldots \alpha_n\}$

3. for each segment $\omega_i$

   compute the link weight $\mu(L_i, \Gamma)$

   compute the image weight $\mu(I_i, \Gamma)$

   compute the theme weight $\mu(T_i, \Gamma)$

   compute the visual feature weight $\mu(V_i, \Gamma)$

   compute the freshness weight $\mu(F_i, \Gamma)$

   sum up the weight components

4. $\Psi$ =Sort (segments, weight)

5. Fetch the template $\vec{\lambda}$

6. Select the top n segments from $\Psi$

7. Render the segments using template $\vec{\lambda}$

8. Place the remaining segments in Segment buffer

End

The MORPES algorithm receive the source page and profile bag as input and renders the mobile version of the page with selected segments as output.

The page is rendered as a sequence of shots. Each shot consists of a collection of segments. The first shot of the mobile version would include those segments for which the segment evaluator has returned a high score. The successive shots would consist of segments in the decreasing order of their weights. The segment buffer consists of those segments which were not displayed in first shot. The successive shots would read from the segment buffer to display the contents. The algorithm can be proved using simple induction technique.

## 4. EXPERIMENTS AND RESULT ANALYSIS

This section explores the experimentation and results associated with the proposed MORPES model for rendering of web pages in mobile devices. The prototype implementation is done with the software stack including Ubuntu Linux , Apache, MySql and PHP at the proxy server. The mobile device is provided with a simple client application to render the contents structured by the MORPES model at the proxy server. For client side scripting JavaScript is used.

With respect to the hardware, a Core i5 processor system with 3 GHz of speed, 8 GB of RAM is used. The internet connection used in the experimental setup is a 128 Mbps leased line.

The experiments were conducted in various sessions. The result data of the experiments are tabulated in Table 1.

Table 1. Experimental Results of the MORPES model

| Session ID | MSC | MSFS | MPSC |
| --- | --- | --- | --- |
| 1 | 25.21 | 5.3 | 5.1 |
| 2 | 16.15 | 4.2 | 7.1 |
| 3 | 18.35 | 2.3 | 4.2 |
| 4 | 22.37 | 1.3 | 5.4 |
| 5 | 20.11 | 4.5 | 3.2 |
| 6 | 23.38 | 5.2 | 6.2 |
| 7 | 21.45 | 3.3 | 6.5 |
| 8 | 22.32 | 5.5 | 7.3 |
| 9 | 18.64 | 6.1 | 5.2 |
| 10 | 22.32 | 5.3 | 1.7 |
| 11 | 16.15 | 1.2 | 1.8 |
| 12 | 17.58 | 4.3 | 3.5 |
| 13 | 17.65 | 3.5 | 1.2 |
| 14 | 13.21 | 4.3 | 3.3 |
| 15 | 12.10 | 4.6 | 3.1 |

In Table 1 MSC stands for mean segment count which indicates the mean of the number of segments in that session. MSFS stands for Mean of Segments in First Shot, and MPSC stands for mean of page shot count. In Fig 2, the chart depicts the comparison of various segment parameters while rendering the web page contents on mobile devices.

It was observed from the experiments across the sessions that the mean of segments at the first shot was 4.06 where as the mean of segments was 19.13. The segments which carry the highest score with respect to the user's informational requirements only are displayed in the first shot there by giving a clear view of the data that user might be interested in.

From the experiments conducted it is also observed that users were able to find their needy contents in the first few shots of the page itself rather than going deep through the segment buffer. This core idea solves the problem of providing the information relevant to the user in the best possible way with the available screen size.

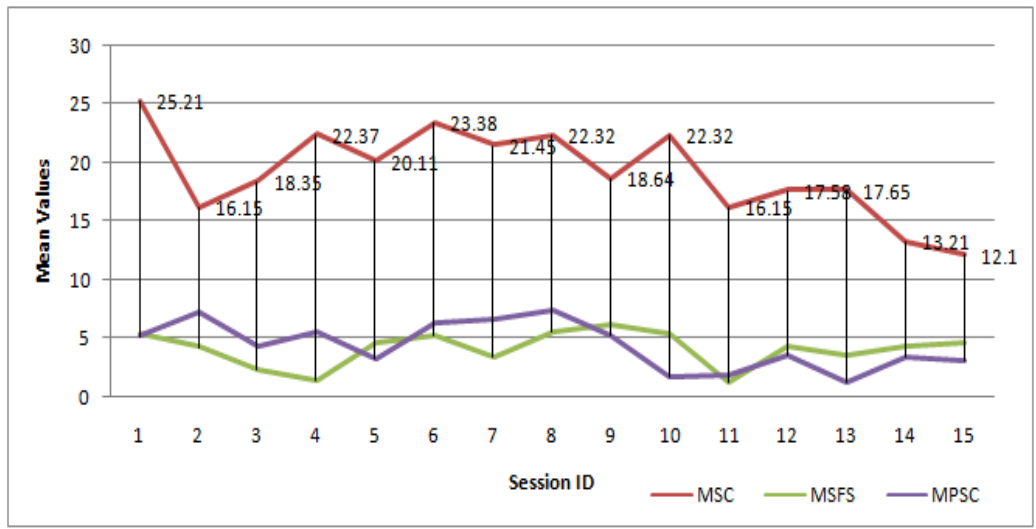

Figure 5. Comparison of Various Segment Parameters

## 5. CONCLUSIONS AND FUTURE DIRECTIONS

The proposed MORPES model which facilitates in rendering of contents on the mobile device, leads to following conclusions:

- The problem of occupying the contents of web pages which are originally designed for personal computers, into mobile devices with smaller screen, is effectively addressed by restructuring the page contents and providing it to the user in multiple shots.
- As the proposed model incorporates the personalized evaluation of segments, the first shot of the page presented to the user is tailored in such a way that the segments with maximum score according to the user's requirements would get a place in the first shot.
- By downsizing the amount of content to be rendered in the first shot using the MORPES approach makes the user to locate the contents in a smaller screen area much easier.

The future directions for the MORPES model include the following:
- The model can be made more effective by introducing a relevance feedback mechanism regarding the segments presented to the user.
- After the incorporation of relevance feedback mechanism, the learning based enhancement can be applied to the model based on soft computing techniques.

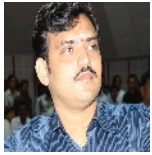
K.S.Kuppusamy is an Assistant Professor at Department of Computer Science, School of Engineering and Technology, Pondicherry University, Pondicherry, India. He has obtained his Masters degree in Computer Science and Information Technology from Madurai Kamaraj University. He is currently pursuing his Ph.D in the field of Intelligent Information Management. His research interest includes Web Search Engines, Semantic Web. He has made more than ten peer reviewed international publications.

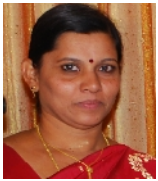
G. Aghila is a Professor at Department of Computer Science, School of Engineering and Technology, Pondicherry University, Pondicherry, India. She has got a total of 22 years of teaching experience. She has received her M.E (Computer Science and Engineering) and Ph.D. from Anna University, Chennai, India. She has published more than 55 research papers in web crawlers, ontology based information retrieval. She is currently a supervisor guiding 8 Ph.D. scholars. She was in receipt of Schrneiger award. She is an expert in ontology development. Her area of interest includes Intelligent Information Management, artificial intelligence, text mining and semantic web technologies.